\documentstyle[11pt]{article}

\def\fspl               {\pl\!\!\!/\,}

\def\fsh                {h\!\!\!/\,}
\def\fsk                {k\!\!\!/\,}

\def\fsA                {A\!\!\!/\,}

\def\fsD                {D\!\!\!\!/\,}

\def\ZZ                 {Z\!\!\!Z\,}

\def\al           {\alpha}
\def\be           {\beta}

\def\de           {\delta}
\def\ep           {\epsilon}

\def\ga           {\gamma}

\def\la           {\lambda}
\def\om           {\omega}

\def\si           {\sigma}

\def\th           {\theta}

\def\De           {\Delta}
\def\Ga           {\Gamma}
\def\La           {\Lambda}

\def\Pi           {\Pi}

\def\non          {\nonumber}

\def\ha           {\mbox{$\frac{1}{2}$}}

\def\pl           {\partial}

\def\d        {d}

\def\bfk        {\mbox{\boldmath $k$}}

\def\spa          {\ \ \ }
\def\mand         {\spa\mbox{and}\spa}

\def\mfor         {\spa\mbox{for}\spa}

\def\mwhere       {\spa\mbox{where}\spa}

\def\mimp         {\spa\mbox{$\Rightarrow$}\spa}

\def\sgn          {\mbox{sgn}}

\def\Tr           {\mbox{\rm Tr}\,}
\def\tr		  {\mbox{\rm tr}\,}
\def\cd           {{\cdot}}

\newcounter {sect}
\def\thesect {\Roman{sect}}
\def\sect#1{
\refstepcounter{sect}
\begin{center}
{\bf \thesect: #1}
\end{center}}

\scrollmode

\begin{document}

\pagenumbering{arabic}
\hfill hep-th/9710088

\vspace{3cm}
\begin{center}
{\Large{\bf Induced Chern-Simons terms}}

\vspace{1cm}
\renewcommand{\thefootnote}{\fnsymbol{footnote}}
Jim McCarthy and Andy
Wilkins\footnote{awilkins,jmccarth@physics.adelaide.edu.au} 
\setcounter{footnote}{0}

{\em Department of Physics and Mathematical Physics \\
University of Adelaide \\
Adelaide 5005 \\
Australia}

(October 10, 1997)

\vspace{5mm}

\begin{minipage}{14cm}
We examine the claim that the effective action of
four-dimensional SU(2)$_{L}$ gauge theory at high and low temperature contains
a three-dimensional Chern-Simons term with coefficient being the chemical
potential for baryon number.  We perform calculations in a
two-dimensional toy model and find that the existence of the term is 
rather subtle.

11.10 Wx, 98.80 Cq
\end{minipage}
\end{center}

\newpage

\sect{INTRODUCTION}

Consider the the 4-dimensional Euclidean SU(2)$_{L}$ gauge 
theory at finite temperature $T = 1/\be$, described by
\begin{equation}
S = \int_{0}^{\be}\d\tau\int\d^{3}x \left(-\ha \tr F^{2} +
\bar{\psi}_{L}\fsD\psi_{L}\right)
\ .
\end{equation}
There are an even number of massless
left-handed fermions to avoid the global SU(2) anomaly~\cite{witten},
and the real chemical potential $\mu$ for the
particle-number charge $B_{L}$ is non zero,
\begin{equation}
B_{L} = \int\d^{3}x\ \bar{\psi}_{L}\ga^{0}\psi_{L} \ .
\end{equation}
It has been suggested by Redlich and
Wijewardhana~\cite{red}, Tsokos~\cite{tso} and Rutherford~\cite{rut},
that --- at both high and low
temperature --- the effective 
action obtained by integrating out the fermions contains a term
reminiscent of the 3-dimensional Chern-Simons 
term with coefficient $\mu$,
\begin{equation}
S_{\mbox{\scriptsize{eff}}} = \mu\int_{0}^{\be}{\d \tau}\int\d^{3}x\
\ep_{ijk}\tr\left(A_{i}\pl_{j}A_{k} - \mbox{$\frac{2}{3}$} g 
A_{i}A_{j}A_{k}\right) + \ldots \ .
\label{eqn.3.dcs}
\end{equation}

This model been used~\cite{dia,lee} to describe baryogenesis by weak
interactions at temperatures around the weak scale in the early
universe. The authors note 
that because of 
the U(1) anomaly, $B_{L}$ is only quasi-conserved.  Then, when the
gauge configurations tunnel 
{}from one vacuum sector to another, baryons will be
created or destroyed.  Because $\mu$ is real, the `Chern-Simons' term
in Eq.~(\ref{eqn.3.dcs}) is not gauge invariant, and so breaks the degeneracy
of the topological vacua.  Thus the system would
be biased to `fall' in one particular direction
resulting in more baryons being created than antibaryons.

Let
us now present a calculation that produces no
`Chern-Simons'  
term at low temperature.  We use Pauli-Villars regularisation which is
manifestly gauge invariant.
Since $\mu$ is real we are only interested in the real 
part of the effective action, $\log\det\fsD\fsD^{\dagger}$.  This means
the model can be `vectorised'~\cite{red,rut,dia}, by adding
$\bar{\psi}_{R}\fsD^{\dagger}\psi_{R}$, 
to obtain a theory of Dirac fermions with an axial quasi-conserved charge
\begin{equation}
S = \int \bar{\psi}(\fspl + ig\fsA^{a}T^{a} +
\mu\ga^{0}\ga^{5})\psi \ .
\label{equ.vec.4d}
\end{equation}
The coefficient of $\mu
A_{\la}^{a}A_{\de}^{a}$ in the `Chern-Simons' term is
\begin{equation}
\Ga^{\la\de 0}(p,M,T)
= \int_{k}\tr
\ga^{\la}\De(k,M)\ga^{0}\ga^{5}
\De(k,M)\ga^{\de}\De(k+p,M) \ .
\end{equation}
Here $\De(k,M)$ is the propagator of a Dirac fermion with mass $M$ and
the integral over momentum space is
$\int_{k} = 
\be^{-1}\sum_{n}\d^{3}\bfk$ for nonzero temperature.
Following~\cite{red,rut} we add a mass $m$ for the fermions at low
temperature.
Expanding the 
denominator in powers of $(2k\cd p + p^{2})(k^{2} + M^{2})^{-1}$ yields
\begin{equation}
\Ga^{\la\de 0}(p,M,T) = C\ep^{0\la\de\al}p^{\al} + O(p^{2}/M) \ .
\end{equation}
Since $C$ is mass independent, Pauli-Villars regularisation will
yield, in apparent contradiction to~\cite{red,tso,rut},
\begin{equation}
\Ga^{\la\de 0}_{\mbox{\scriptsize{PV}}}(p,m,T\!\sim\!0) 
 \equiv \lim_{M\rightarrow \infty}\left[
\Ga^{\la\de 0}(p,m,T\!\sim\!0) -
\Ga^{\la\de 0}(p,M,T\!\sim\!0) \right]
=  0 + O(m^{-1}) \ .
\end{equation}

It is tempting to invoke gauge invariance in order to rule out the
appearance of the `Chern-Simons' term.  However, this is too naive,
because --- although the term is not gauge invariant by  
itself --- it is still possible that the entire effective action may be
invariant~\cite{rut,dun,des}.   In later sections we shall 
present simple examples of this phenomena.

In light of the apparent contradiction of Pauli-Villars regularisation
with the 
results of~\cite{red,tso,rut}, and the subtlety of gauge invariance, we
feel that the problem needs 
more study.  Fortunately, there is a related model
in two dimensions in which further calculations can be made more simply.

\sect{THE TOY MODEL}

We work in a flat two dimensional  (2D) Euclidean space ${\cal M}$
with coordinates 
$(\tau,x)$ where $0 \leq \tau \leq \be$.  Our gamma matrices are
Hermitian and satisfy
\begin{equation}
\left[ \ga^{\mu},\ga^{\nu} \right]_{+} = 2\de^{\mu\nu}
\mand
\ga_{5}= -i\ga_{0}\ga_{1} \ .
\end{equation}
The 2D equivalent of the vectorised theory of
Eq.~(\ref{equ.vec.4d}) is
\begin{equation}
Z[A,\mu,\bar{\eta},\eta] = \int \left[ \d\bar{\psi}\d \psi
\right] e^{-S - \int \bar{\eta}\psi - \bar{\psi}\eta} \ ,
\label{two.d.toy}
\end{equation}
with
\begin{equation}
S = \int_{\cal M} \bar{\psi}\fsD\psi \mand \fsD = \fspl + m +
 \mu\ga^{0}\ga^{5} + ie\fsA \ .
\end{equation}
A mass term has been included for generality at this point.  We shall
see later on that it infrared (IR) regulates the theory at zero
temperature.  The chemical potential $\mu$ for the Hermitian axial charge
$Q_{5} =\int\bar{\psi}\ga^{0}\ga^{5}\psi$ is real.  One can check this
through a derivation of the path-integral representation of the
partition function.\footnote{The final part of this process is to
express the action derived in terms of relativistic fields in
Euclidean space.  It can be shown
that with the choice $\bar{\psi} = \psi^{\dagger}\ga^{5}$, the path
integral given in Eq.~(\ref{two.d.toy}) calculates the partition
function, thereby confirming the recent work of Waldron
et al.\ \cite{wal}.}

The U(1) gauge
transformations are 
\begin{eqnarray}
A_{\mu} & \rightarrow & A_{\mu} -ie^{-1}e^{i\th}\pl_{\mu} e^{-i\th}
 \non \\
\psi & \rightarrow & e^{i\th}\psi \ .
\label{gauge.trans}
\end{eqnarray}
A gauge transformation is called `small' when $\th$ is well defined on
${\cal M}$, while if only
$e^{i\th}$ is well defined (but not $\th$ itself) the transformation
is called `large'.  An example of a large gauge transformation is
\begin{equation}
\th(x,\tau) = 2\pi \tilde{N}\tau/\be \ , \mfor \tilde{N} \in Z\!\!\!Z
\ .
\end{equation}
This shifts $A_{0}$ by a constant
\begin{equation}
A_{0} \rightarrow A_{0} -2\pi \tilde{N}/e\be \ .
\end{equation}

The `Chern-Simons' term in this context is
\begin{equation}
\mu\int_{\cal M} A^{1} \ .
\end{equation}
Let us first present some perturbative calculations that suggest that
this term does not appear in the effective action.  Then we will study
the effective action nonperturbatively.

\sect{PERTURBATIVE RESULTS}

Since $\mu$ is constant, it
is efficient to put it into the propagator 
\begin{equation}
\De(k) = \frac{1}{i\fsk + m + \mu\ga^{0}\ga^{5}} = \frac{1}{i\fsk + m
- i\mu\ga^{1}} \ .
\end{equation}
The second equality holds in two dimensions because of the identity
$\ga^{\la}\ga^{5} = -i\ep^{\la\de}\ga^{\de}$ and shows that a constant
$\mu$ simply shifts the momentum in the loop.  Expanding the path
integral in powers of $A$ we find the coefficient of the
linear term is the superficially linearly divergent one-point
function 
\begin{equation}
\Ga^{\la}(m,T,\mu) =
%
 \int_{k}\tr e\ga^{\la} \frac{m -
i\tilde{\fsk}}{\tilde{k}^{2} + 
m^{2}} \mwhere \tilde{k}_{1} \equiv k_{1} - \mu \ .
\label{equn.1.pt.f}
\end{equation}
To regulate this expression we will use Pauli-Villars regularisation in
which a massive spinor,
$\chi$, is added into the path integral\footnote{In principle two
spinors are  
needed, however, this is an unnecessary notational complication.}
\begin{equation}
Z = \lim_{M\rightarrow \infty} \int [\d\bar{\psi}\d\psi
\d\bar{\chi}\d\chi
]
e^{-S(\bar{\psi},\psi,A,m) + S(\bar{\chi},\chi,A,M)} \ .
\end{equation}
This is manifestly gauge invariant and, in the usual fashion, gives
\begin{equation}
\Ga^{\la}_{\mbox{\scriptsize{PV}}}(m) \equiv \lim_{M\rightarrow\infty}
\left[\Ga^{\la}(m) - \Ga^{\la}(M)\right] \ .
\label{eqn.pv.g}
\end{equation}
Since the momentum integral is now finite we can shift away all
dependence on $\mu$.  It is possible to go further and explicitly
calculate each 
separate term on the RHS of Eq.~(\ref{eqn.pv.g}).  The mass term
in the numerator of Eq.~(\ref{equn.1.pt.f}) gets killed 
by $\tr\ga^{\mu} = 0$.  When $\la=0$ symmetric summation (or
integration) gives $\Ga^{0}(m,T) = 0$.  For $\la=1$ the answer
obtained depends on the order of integration.  Performing the
$k^{1}$ integral first gives
\begin{eqnarray}
\Ga^{1}(m,T) & = & e\int_{k_{0}}\int_{-\La - \mu}^{\La -
\mu}\d\tilde{k}_{1}\frac{\tilde{k}_{1}}{\tilde{k}_{1}^{2} + m^{2} +
k_{0}^{2}} 
\non \\
& \stackrel{\La \rightarrow \infty}{\longrightarrow} & \int_{k_{0}} 0
\spa  =\spa  0
\ .
\end{eqnarray}
However, performing the $k_{0}$ summation first yields
\begin{eqnarray}
\Ga^{1}(m,T) & = & \be^{2}e\int \d k_{1}
\frac{\tilde{k}_{1}}{\be\sqrt{\tilde{k}_{1}^{2} + m^{2}}}
\,\pi\mbox{tanh}
\left(\pi\be\sqrt{\tilde{k}_{1}^{2} + m^{2}}\right) 
\non \\
& = &  2e\pi\mu \ .
\end{eqnarray}
The same result is obtained at zero temperature.  However, all answers
are mass independent, so Pauli-Villars
regularisation yields
\begin{equation}
\Ga^{\la}_{\mbox{\scriptsize{PV}}}(m,T) = 0 \spa\spa
\forall m,T \ .
\end{equation}

An alternative treatment is not to put $\mu$ into the propagator, but
to expand the path integral in powers of both $\mu$ and $A$.  
The correlation-function of interest is the logarithmically divergent 
two-point function
\begin{equation}
\Ga^{\la 0}(m,T)
= \int_{k}\tr \frac{m - i\fsk}{k^{2} +
m^{2}}\ga^{0}\ga^{5} 
 \frac{m - i\fsk}{k^{2} + m^{2}}ie\ga^{\la}
\ .
\label{two.pt.fcn}
\end{equation}
This method has the advantage that we can easily make $\mu$
nonconstant.  The
momentum $p$, flowing into the associated Feynman diagram will then be
nonzero, and only 
after calculating will we set $p =0$.  With nonzero $p$,
Adler's regularisation-independent method~\cite{adler} can be 
applied.  At zero temperature, the most general 
expression with the correct Lorentz structure and parity is
\begin{equation}
\Ga^{\la\de}(p,m,T\!=\!0) = Y(p^{2},m^{2})\ep^{\la\de}
+ Z(p^{2},m^{2})p_{\si}\ep^{\si\left(\la\right.}p^{\left.\de\right)} \
.
\label{equ.ad.zero.t}
\end{equation}
The parentheses indicate symmetrisation.  Gauge invariance implies
\begin{equation}
p_{\la}\Ga^{\la\de} = 0 \mimp p_{1}\Ga^{10} = p_{0}\Ga^{00} \mimp
Y = -\ha p^{2}Z \ .
\end{equation}
But $Z$ is finite so we can calculate it.  For the massive case we
find $Z \propto 
m^{-2} + O(p^{2})$.  Then setting $p^{2}=0$ gives
\begin{equation}
Y = 0 \mimp \Ga^{\la\de}(m\!\neq\! 0,T\!=\! 0) = 0 \ .
\end{equation}
However, for $m=0$ we obtain
\begin{equation}
\Ga^{10}(p,m\!=\! 0,T\!=\! 0) =\frac{2e\pi p_{0}^{2}}{p_{0}^{2} + p_{1}^{2}}
\ .
\label{ga10,ad}
\end{equation}
Interestingly, this is ambiguous in the zero-momentum limit
\begin{equation}
\Ga^{10}(m\!=\! 0,T\!=\! 0) \rightarrow
\left\{
\begin{array}{ll}
0 & p_{0}\rightarrow 0 \ \mbox{then}\  p_{1} \rightarrow 0
\\
2e\pi & p_{1} \rightarrow 0 \ \mbox{then}\  p_{0} \rightarrow 0 \ .
\end{array} \right. 
\label{ad.amb}
\end{equation}

We attribute this to the IR divergence contained in the two-point
function of Eq.~(\ref{two.pt.fcn}) for $M=0$ and $T=0$.  We find a 
similar problem when naively applying Pauli-Villars regularisation 
at zero temperature.  Namely, after taking
the trace over gamma matrices,
\begin{eqnarray}
\Ga^{\la\de}(M\! \neq \! 0,T\!=\! 0) & = &
ieM^{2}\tr\ga^{\de}\ga^{5}\ga^{\la} \int_{k} (k^{2} +
M^{2})^{-2}
\non \\
& = & -2e\pi\ep^{\la\de} \ ,
\end{eqnarray}
while
\begin{equation}
\Ga^{\la\de}(M \! = \! 0,T\!=\! 0) = 0 \ .
\end{equation}
This implies, in contradiction to the null result obtained using the one-point
function,
\begin{equation}
\Ga_{\mbox{\scriptsize{PV}}}^{1 0}(m,T\!=\! 0) =
\left\{
\begin{array}{ll}
0 & m\neq 0
\\
2e\pi & m=0 \ .
\end{array}
\right.
\end{equation}
However, this occurs only because the IR divergence has made the
result somewhat arbitrary.  In this situation a natural prescription
is to  define the massless theory as the limit of the massive one:
\begin{equation}
\Ga^{10}_{\mbox{\scriptsize{PV}}}(m,T\!=\! 0) = 0 \spa\spa \forall m \ .
\end{equation}

At nonzero temperature there is no IR problem because $k_{0}$ is
never zero.  Pauli-Villars regularisation gives zero in agreement with
the one-point function.  The Adler argument is more
complicated because the heat 
bath breaks Lorentz invariance and so $\Ga^{\la\de}$ can depend on the
normal vector in the $p_{0}$ direction.  It turns out~\cite{steve},
that $\Ga^{10}$ has the same form as Eq.~(\ref{ga10,ad}).
However, this time $p_{0}$ is quantised, which means it can't be taken
to zero smoothly.  We argue that this implies that $p_{0}$ must be set
to zero from the very start, and so the top limit in
Eq.~(\ref{ad.amb}) is the correct one.

\sect{NONPERTURBATIVE RESULTS}

The partition function can also be calculated directly to all orders
in $\mu$ by functional methods.\footnote{We are interested in the
trivial sector of 
the model.  The effective action when the gauge field is in a nontrivial
winding sector is also well known~\cite{non,sac}.  Nontrivial sectors may
be of interest when studying baryogenesis in the early universe.}
To make the
eigenvalue problem well-defined, ${\cal M}$ is chosen to be the torus
with $0\leq\tau\leq\be$ and $0\leq x\leq R$.  Here we can make the Hodge
decomposition on the background gauge field 
\begin{equation}
A_{\mu} =\mbox{$\frac{1}{e}$} \pl_{\mu}\si +
\mbox{$\frac{1}{e}$}\ep_{\mu\nu}\pl_{\nu}\rho + h_{\mu} \ .
\label{equ.gauge.f}
\end{equation}
The fields $\si$ and $\rho$ are well defined on ${\cal M}$ and $h_{\mu}$ is
constant.
Our case differs from the Schwinger model~\cite{sch} on the torus only by the
$\mu$ term.  However, using the identity $\ga^{0}\ga^{5} = -i\ga^{1}$
we can shift the $\mu$ into $h_{1}$.  The form of the generating
functional is well known~\cite{jac}
\begin{equation}
Z[A,\bar{\eta},\eta] = \exp \left( \int
\bar{\eta}e^{-i\si-\ga^{5}\rho}\De_{0}e^{i\si \ga^{5}\rho}\eta
+  \frac{1}{2\pi}\int\rho\Box\rho\right)\,
\det\fsD_{0} \ .
\label{e.with.r.and.z}
\end{equation}
Here $\fsD_{0} = \fspl + ie\fsh - i\mu\ga^{1}$ and has
associated propagator $\De_{0}$.  The determinant of this operator can
be calculated 
using zeta-function regularisation.  The result can be written in
terms of a theta function and Dedekind's eta 
function~\cite{alv,sac}  
\begin{eqnarray}
\det \fsD_{0} & = & \left| \frac{1}{\eta(iR/\be)}
\Theta\left[\begin{array}{c}\th \\ \phi
\end{array}\right](0,iR/\be)\right|^{2}  
\non \\
& = & \left| q^{1/24} \prod_{m=1}^{\infty}(1-q^{m})
\sum_{n\in\ZZ}q^{\ha(n+\th)^{2}}e^{2\pi i(n+\th)\phi} \right|^{2} 
\ .
\label{trivial.det}
\end{eqnarray}
In this formula $\th = -\be eh^{0}/2\pi$ and  $\phi = \ha +
\mbox{$\frac{R(eh^{1}-\mu)}{2\pi}$}$ and the parameter $q=e^{-2\pi
R/\be}$. 

The partition function is clearly
invariant under small gauge transformations since
$e^{i\si}\eta$ and its conjugate are invariant.  It is also
invariant under large gauge transformations in the $x$ and $\tau$
directions 
\begin{eqnarray}
\mbox{`$x$' direction:} &&\!\!\!\!
 \de h^{1} = \mbox{$\frac{2\pi
\tilde{N}}{eR}$} \mand \bar{\eta} \rightarrow \bar{\eta}e^{2\pi i
\tilde{N}x/R}
\ ,\spa \non \\
\mbox{`$\tau$' direction:} && \!\!\!\! \de h^{0} = \mbox{$\frac{2\pi
\tilde{N}}{e\be}$} \mand \bar{\eta} \rightarrow \bar{\eta}e^{2\pi i
\tilde{N}\tau/\be} \ .
\end{eqnarray}
The first transformation changes the summand in
Eq.~(\ref{trivial.det}) by  
a phase which is then canceled by the mod-squared.  The second
transformation can be soaked up by relabeling the index of summation.

Let us study the partition function as we take the cylindrical limit.
The 
determinant~(\ref{trivial.det}) of $\fsD_{0}$ obtained by
zeta-function regularisation is 
nonlocal in the gauge field.  Also, each term in the expansion of the
effective action $S_{\mbox{\scriptsize{eff}}} = \log\det\fsD_{0}$ in
powers of $h^{\la} = 
\frac{1}{R\be}\int 
A^{\la}$ is not gauge invariant.  For example, at large $R$ (the limit
to the cylinder) or small $\be$ (high temperature), the parameter $q$ is
small.  Then we can expand for $\th=0$
\begin{equation}
S_{\mbox{\scriptsize{eff}}} =
 8\sqrt{q}\frac{R}{\be}e\mu\int
A^{1} + \ldots \ ,
\label{eqn.small.q}
\end{equation}
where, in the last equality, the `Chern-Simons' term has been
extracted.  The term by itself 
is not gauge invariant.  In the appendix we study the one
dimensional analogue, $\det\fsD$ on the circle.  Once again
zeta-function regularisation results in 
a nonlocal but gauge-invariant result.  Each term in the expansion in
powers of the gauge field is not gauge invariant.  We also study the
limit to the line.  One would not expect the 
limit to depend upon whether the boundary conditions on the circle were
initially periodic or antiperiodic.  The only subtlety is
that one has to be careful
with IR divergences (zeromodes).  In the 2D model there
are no IR 
problems because the fermions are antiperiodic along the time
direction.  Thus, by setting $q=0$ in Eq.~(\ref{eqn.small.q}), we
see that there 
is no induced Chern-Simons term on the cylinder according to
zeta-function regularisation.

\sect{CONCLUSIONS}

The effective action of the
2D toy model of baryogenesis has been calculated in various ways.
Because the chemical 
potential is real, the 
Chern-Simons-type term that has been proposed to appear in the
effective action is not gauge invariant.  As
we have seen in one and two dimensions, this does not rule out its
appearance in the effective action.  However, all our gauge-invariant
calculations at nonzero temperature gave no Chern-Simons term.  It
was only for the massless theory at zero temperature that there was
any chance of getting a term.  This was attributed to an ambiguity
brought about through an IR divergence.

How then, did other authors~\cite{red} obtain a nonzero result?  The
regularisation scheme was to subtract off the zero-temperature,
zero-$\mu$ result.  Let us perform the same calculation in 2D. 
The one-point function of
Eq.~(\ref{equn.1.pt.f}) can be written in the form
\begin{equation}
\Ga^{1}(m,T,\mu) = \int \d k_{1} \oint_{C} \frac{\d z}{2\pi i}
\left(\frac{k_{1} - \mu}{-z^{2} + (k_{1} - \mu)^{2} + m^{2}}\right)
\tanh \ha\be z  \ ,
\end{equation}
where the contour of integration is shown in Fig. 1(a).
Using partial fractions and expressing $\tanh$ in terms of
exponentials leads to
\begin{eqnarray}
\Ga^{1}(m,T,\mu) & = & \int \d k_{1} \frac{k_{1} - \mu}{\om}
\left[
- \oint_{\overline{C}_{+}} \frac{\d z}{2\pi i} \left( \frac{1}{z+w} -
\frac{1}{z-w} \right) \frac{1}{1 + e^{\be z}} \right.
\non \\
&& \spa\spa\spa
-\left. \oint_{\overline{C}_{-}} \frac{\d z}{2\pi i} \left( \frac{1}{z+w} -
\frac{1}{z-w} \right) \frac{1}{1 + e^{-\be z}} 
+ \int_{\overline{C}_{0}} \frac{\d z}{2\pi i} \left( \frac{1}{z+w} -
\frac{1}{z-w} \right) \right] \ ,
\end{eqnarray}
where $\om = \sqrt{(k_{1}-\mu)^{2} + m^{2}}$ and the various contours
are shown in Fig. 1(b).
Evaluating these integrals leads to
\begin{equation}
\Ga^{1}(m,T,\mu) = 2e\pi\mu + \Ga^{1}(m,0,0) \ .
\end{equation}
Thus, if we follow~\cite{red} and regulate by
subtracting off the zero-temperature, zero-$\mu$ 
result, we will obtain a `Chern-Simons' term.  This is in contrast to
Pauli-Villars regularisation which gave no `Chern-Simons' term.

One might try to justify this procedure by casting it into a
Pauli-Villars-like form
\begin{equation}
Z = \lim_{M\rightarrow\infty}\int [\d\bar{\psi}\d\psi
\d\bar{\chi}\d\chi
]
\exp\left[-S(\bar{\psi},\psi,A,m,T,\mu)
+ S(\bar{\chi},\chi,A,M,T=0,\mu=0)\right] \ . 
\label{unusual.pv}
\end{equation}
In the second action the spinor fields $\chi$ are defined over the
plane.  The gauge
field must be the same in both actions.   Presumably it is extended
periodically to the plane in the second action.  The second action
also has no axial charge.  A standard argument
shows that there are no new divergences introduced by insertions of
the charge of a conserved current.  In the present case, $Q_5$ is the
charge of an
anomalous current, so this argument must be re-examined.  Clearly it
is somewhat uncertain as to whether this scheme can be implemented as
a gauge-invariant regularisation to all orders in perturbation
theory.  In contrast, the
regularisation schemes used in 
this paper are gauge invariant and implementable to all orders.
If the unusual regularisation scheme in Eq.~(\ref{unusual.pv}) can be
implemented then it amounts to a definition of 
the theory, and it would be interesting to re-examine the
cosmological models using it to see whether the `Chern-Simons' term
arises in their effective 
description.  Using zeta function regularisation, the effective action
for gauge fields in nontrivial winding sectors has also been
calculated~\cite{non,sac}.  It would be of interest to
calculate matrix elements corresponding to baryogenesis in the early
universe with this action.

\begin{center}
ACKNOWLEDGMENTS
\end{center}

We would like to thank Steve Poletti for the initial impetus behind
this project and many useful discussions.  We are grateful to Stanley
Deser for reading and commenting on this manuscript.


\begin{center}
APPENDIX: DETERMINANT ON A ONE-DIMENSIONAL MANIFOLD
\end{center}

A nonperturbative result for the partition
function on the torus has been presented.  The effective action was
nonlocal and the expansion in small $A$ naively
looked gauge variant.  The one-dimensional theory has these properties
too.  It also provides us with a testing ground to check for
nontrivialities in the torus $\rightarrow$ cylinder limit.  Start
with the operator
\begin{displaymath}
D = i\pl + eA(t) + iM \ ,
\end{displaymath}
where $-\pi R \leq t \leq \pi R$.  We have included a mass term $iM$
for generality, and it will serve to IR regulate the theory.
On the circle the eigenvectors are
\begin{displaymath}
\psi_{\la} = \exp \left[ i\left(\la t - e\int^{t}A \right) - M
t\right] \ .
\end{displaymath}
The boundary conditions then imply $\la_{n} = {\cal A} + (n/R)$ where
\begin{displaymath}
{\cal A} \equiv \left\{
\begin{array}{ll} 
 \frac{e}{2\pi R}\int A - iM
& \spa\mbox{periodic}
\\
\frac{1}{2R} + \frac{e}{2\pi R}\int A - iM
& \spa\mbox{antiperiodic} \ .
\end{array}
\right.
\end{displaymath}
If $M\neq 0$ there are no zeromodes, however, if $M=0$ there is a
possibility of one zeromode depending on the value of $\int A$.
The product of eigenvalues needs regularisation.  A
non-gauge-invariant way to proceed is to calculate $\det D(i\pl + 
iM)^{-1}$.  This leads to a sine in the periodic case and a cosine for
antiperiodic boundary conditions.  Alternately, zeta-function
regularisation is gauge invariant, and results in (for values of the
Riemann zeta 
function see Ref.~\cite[\S9.53]{GR})
\begin{eqnarray}
\det D & = & \exp - \frac{\d}{\d s}\sum_{n} \left.\left(\frac{n}{R} +
{\cal A} \right)^{-s} \right|_{s=0}
\ , \non \\
& = & 1- e^{-2\pi i {\cal A} R} \ .
\non
\end{eqnarray}

Consider the antiperiodic massless theory.  Expanding the effective
action in powers of $A$ gives
\begin{displaymath}
S_{\mbox{\scriptsize{eff}}} = \log 2 - \ha e i\int A + O(A^{2}) \ .
\end{displaymath}
Although the whole effective action is gauge invariant, this term is
only invariant under $A \rightarrow A - 2\pi\tilde{N}/eR$ for even
$\tilde{N}$.  It is
clear that the effective action for the periodic massless 
case does not have an expansion in small $A$.  This is because
there is a zeromode which must be removed
\begin{displaymath}
\det{}'_{\mbox{\scriptsize{periodic}}}D = \frac{1 - e^{-ie\int
A}}{ie\int A} \ .
\end{displaymath}
The same problem crops up in perturbation theory, where we get IR
divergent terms such as
$\sum_{n} \frac{1}{n}$.

The limit to the line of the above result is (mod$2\pi i$)
\begin{displaymath}
\log\det D  \rightarrow
-\pi R(M - |M|) - i\th(-M)e\int A 
+ \left\{
\begin{array}{l}\pi i \\ 0 \end{array}
\right.
\begin{array}{l}\mbox{periodic} \\ \mbox{antiperiodic} \end{array} \ ,
\end{displaymath}
for $ M\neq 0 $, while for $M=0$ the antiperiodic case gives
\begin{displaymath}
\log\det D \rightarrow \log\left( 1+ e^{-ie\int A} \right) \ .
\end{displaymath}
The $M$-dependent normalisation is physically unimportant.  If we had
taken the limit of the massless periodic case without removing the
zeromode, the effective action would not have had an expansion in
small $A$.  It is only when the compact theory is properly IR
regulated that the noncompact effective action can be properly
defined.  In our
2D example, the antiperiodicity over the time direction at
nonzero temperature will provide the necessary IR regulator.

Let's
compare this with the expression obtained from $\det D(i\pl + 
iM)^{-1}$.  The Green's function for $i\pl + iM$ with $M\neq 0$ is
\begin{eqnarray}
G(x-y) & = & \int \frac{\d k}{2\pi} \frac{e^{ik(x-y)}}{-k + iM} 
\non \\
&= &
\left\{ 
\begin{array}{ll}
ie^{-M(x-y)}\left[\th(M)\th(x-y)
-\th(-M)\th(y-x) \right] & \mbox{for }
x-y \neq 0
\\
-\ha i \sgn M & \mbox{for } x-y= 0 \ .
\end{array}
\right.
\non
\end{eqnarray}
where $\th$ is a step function.  Expanding the effective action in
powers of $A$, the step functions kill all terms but the linear one,
resulting in
\begin{displaymath}
\det D(i\pl+ iM)^{-1} = \exp \ha i\sgn M\int_{-\infty}^{\infty}\d x
A(x) \ .
\end{displaymath}
Because there are no large gauge transformations on the line this is
gauge invariant.  It it differs from the zeta function result
$-i\th(-M)\int A$.  It is well-known that the
imaginary part of the effective action can be defined in many ways
(see~\cite{ball} for a review).

As in the 2D case, zeta function regularisation has
resulted in a nonlocal expression for the effective action.  It is of
interest to see if the derivative expansion, which is local,
feels these non-localities in any way.  To calculate the derivative
expansion we use the heat-kernel method.
This has the disadvantage that only the real part of the effective
action, $\log\det DD^{\dagger}$, can be calculated, because the heat
kernel is then quadratic in derivatives.
However, it has the advantage that at finite $R$ we can apply the
well-known result that the heat kernel is not temperature ($R$)
dependent (see for example~\cite{boschiET}).  Then
\begin{eqnarray}
\log\det DD^{\dagger} & = & \int_{0}^{\infty}\frac{\d \ep}{\ep}\Tr e^{-\ep
DD^{\dagger}}
\non \\
& = & \int_{0}^{\infty}\frac{\d \ep}{\ep}
e^{-\ep M^{2}}\int \frac{\d k}{2\pi}\, e^{ikx}
e^{-\ep(-\pl^{2} +
2iA\pl + (i\pl 
A + A^{2}))}e^{ikx}
\non \\
& = & \int_{0}^{\infty}\frac{\d \ep}{\ep}
\frac{1}{\sqrt{\ep}}e^{-\ep M^{2}}\int \frac{\d k}{2\pi}\,
e^{-k^{2}}
e^{-2\sqrt{\ep}k D_{0} - \ep D_{0}D_{0}}
\non \\
& = & \int_{0}^{\infty}\frac{\d \ep}{\ep}
\frac{1}{\sqrt{4\pi\ep}}e^{-\ep M^{2}} \ ,
\non
\end{eqnarray}
where $D_{0} = i\pl + A$.  The last line follows by expanding the
exponential in powers of $\ep$.  Thus, the real part of the effective
action does not depend on the gauge field $A$.  This does not agree
with the nonlocal zeta-function result.  It is, however, the same as
$\det D(i\pl +  
iM)^{-1}$ on the line.


\newpage

\noindent\begin{center}
\mbox{\parbox{4cm}{$\mbox{}$
\put(10,-52){
\put(0,-0.2){\vector(1,0){80}}
\put(0,0.2){\vector(1,0){80}}
\put(40.2,-30){\line(0,1){22}}
\put(40.2,0){\vector(0,1){30}}
\put(39.8,-30){\line(0,1){22}}
\put(39.8,0){\vector(0,1){30}}
\put(38,-6){\scriptsize{$0$}}
\put(40,0){\circle*{1.5}}
\put(40,0){\oval(8,65)}
\put(44,10){\vector(0,1){4}}
\put(36,-10){\vector(0,-1){4}}
\put(44,16){\scriptsize{$C$}}
\put(35,-50){(a)}
}
}}
\mbox{\parbox{4.3cm}{$\mbox{}$
\put(20,0){
\put(0,0.2){\vector(1,0){80}}
\put(0,-0.2){\vector(1,0){80}}
\put(40.2,-30){\line(0,1){22}}
\put(40.2,0){\vector(0,1){30}}
\put(39.8,-30){\line(0,1){22}}
\put(39.8,0){\vector(0,1){30}}
\put(38,-6){\scriptsize{$0$}}
\put(40,-40){\line(0,1){32}}
\put(40,0){\vector(0,1){43}}
\put(36,-40){\vector(0,1){54}}
\put(36,14){\line(0,1){26}}
\put(44,-40){\vector(0,1){54}}
\put(44,14){\line(0,1){26}}
\put(40,0){\circle*{1.5}}
\put(36,0){\oval(90,80)[l]}
\put(44,0){\oval(90,80)[r]}
\put(-9,-6){\vector(0,-1){4}}
\put(89,-6){\vector(0,-1){4}}
\put(45,12){\scriptsize{$\overline{C}_{+}$}}
\put(23,12){\scriptsize{$\overline{C}_{-}$}}
\put(37,45){\scriptsize{$\overline{C}_{0}$}}
\put(35,-50){(b)}
}
}}
\end{center}

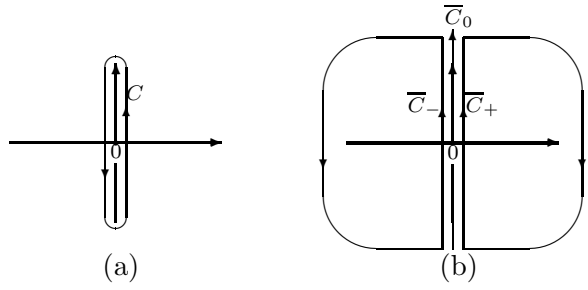
\noindent FIG. 1.  Contours of integration in the $z$-plane.  (a) The
contour $C$
encircles the imaginary axis, and (b) contour $\overline{C}_{0}$ passes up
the imaginary axis and $\overline{C}_{\mbox{\scriptsize{+}}}$
($\overline{C}_{-}$) encircles the RHS (LHS) of the
plane.

\end{document}